\documentstyle[pra,epsf,aps,twocolumn]{revtex}

\begin{document}
\title{{\LARGE Supersymmetry in Stochastic Processes with Higher-Order Time
Derivatives}}
\author{Hagen KLEINERT\ \ \ and\ \ \ Sergei V. SHABANOV \thanks{%
Humboldt fellow; on leave from Laboratory of Theoretical Physics, JINR,
Dubna, Russia.} \thanks{
Email: kleinert@physik.fu-berlin.de; shabanov@physik.fu-berlin.de; URL:
http://www.physik.fu-berlin.de/\~{}kleinert. Phone/Fax:
 0049/30/8383034 }}
\address{Institut f\"ur Theoretische Physik,\\
Freie Universit\"at Berlin, Arnimallee 14, 14195 Berlin, Germany}
\maketitle

\begin{abstract}
A supersymmetric path integral representation is developed for stochastic
processes whose Langevin equation contains any number $N$ of time
derivatives, thus generalizing the Langevin equation with inertia studied by
Kramers, where $N=2$. The supersymmetric action contains $N$ fermion fields
with first-order time derivatives whose path integral is evaluated for
fermionless asymptotic states.
\end{abstract}

{\bf 1}. For a stochastic time-dependent variable $x_{t}$ obeying a
first-order Langevin equation
\begin{equation}
L_{t}[x]\equiv \dot{x}_{t}+F(x_{t})=\eta _{t},  \label{lan}
\end{equation}
driven by a white noise $\eta _{t}$ with $\langle \eta _{t}\rangle =0,\
\langle \eta _{t}\eta _{t^{\prime }}\rangle =\delta _{tt^{\prime }}$, the
correlation functions $\langle x_{t_{1}}\cdots x_{t_{n}}\rangle $ can be
derived from a generating functional
\begin{equation}
Z[J]=\langle e^{i\int dtJx}\rangle ={\int }{\cal D}x\,\Delta
\,e^{-S_{b}+i\int dtJx},  \label{z}
\end{equation}
with an action $S_{b}=\frac{1}{2}\int dt\,L_{t}^{2}$, and a Jacobian $\Delta
={\rm det}\,\delta _{x_{t^{\prime }}}L_{t}$. We denote by $\delta
_{x_{t^{\prime }}}L_{t}$ the functional derivative of $L_{t}[x]$ with
respect to its argument. Explicitly: $\delta _{x_{t^{\prime
}}}L_{t}=[\partial _{t}+F^{\prime }(x_{t})]\delta _{tt^{\prime }}$. The time
variable is written as a subscript to have room for functional arguments
after a symbol. It was pointed out by Parisi and Sourlas \cite{parisi} that
by expressing the Jacobian $\Delta $ as a path integral over Grassmann
variables
\begin{equation}
\Delta ={\int }{\cal D}\bar{c}{\cal D}c\,e^{-S_{f}}  \label{det}
\end{equation}
with a fermionic action
\begin{equation}
S_{f}=\int dtdt^{\prime }\,\bar{c}_{t}\,\delta _{x_{t^{\prime
}}}L_{t}\,c_{t^{\prime }}=\int dt\,\bar{c}_t(\partial _{t}+F^{\prime
})c_t\ ,
\label{ferac}
\end{equation}
the combined
action $S\equiv S_{b}+S_{f}$ becomes invariant under
supersymmetry transformations generated by the nilpotent ($Q^{2}=0$)
operator
\begin{equation}
Q={\int }dt\,(ic\delta _{x}-iL_{t}\delta _{\bar{c}})  \label{qx}
\end{equation}
The supersymmetry implies $QS=0$.

The determinant (\ref{det}) should not be confused with the partition function
for fermions governed by the Hamiltonian
associated with the action (\ref{ferac}). Instead of a trace over
external states
it contains only the vacuum-to-vacuum transition amplitude
for the
imaginary-time interval under consideration.
In the coherent state representation, $\bar{c}_t$
and
$c_t$ are set to zero at the
initial and final
times,
respectively \cite{klauder}.

The path integral (\ref{z}) can also be rewritten in a canonical Hamiltonian
form by introducing an auxiliary Gaussian integral over momentum variables $%
p_t$, and replacing $S_b$ by $S^{H}_b=\int dt\,(p^2_t/2 - ip_tL_t)$. The
generator of supersymmetry for the canonical action is $Q^{H} ={\textstyle%
\int} dt\,(ic_t\delta_{x_t} + p_t\delta_{\bar{c}_t})$. This form has an
important advantage to be used later that it does not depend explicitly on $%
D_t$, so that the above analysis remains valid also for more general colored
noises with an arbitrary correlation function $\langle
\eta_{at}\eta_{bt^{\prime}}\rangle =
(D_{ab})_{tt^{\prime}}\neq \delta_{tt^{\prime}}
$.

Inserting (\ref{det}) into (\ref{z}), the generating functional becomes
\begin{equation}
Z[J] \!=\! \langle e^{i\int dt Jx}\rangle\! =\! {\int} {{\cal D}p} {\cal D}x
{\cal D} \bar{c} {\cal D} c\, \,e^{-S^{H} -S_f+ i\int dt Jx}.  \label{zH}
\end{equation}
This representation makes the stochastic process (\ref{lan}) equivalent to
by supersymmetric quantum mechanical system in imaginary time. In the
supersymmetric formulation of a stochastic process, there exists an infinity
of Ward identities between the correlation functions which can be collected
in the functional relation
\begin{equation}
{\int} {{\cal D}p} {\cal D}x {\cal D}\bar{c}{\cal D}c \,e^{-S^{H}}
Q^H\,\Phi= 0,  \label{wi}
\end{equation}
valid for an arbitrary functional $\Phi\equiv \Phi[p,x,\bar c,c]$. The
Ward identities simplify a perturbative computation of the correlation
functions.

A proof of the equivalence of (\ref{lan}) to (\ref{z}) requires a
regularization of the path integral, most simply by time slicing. This is
not unique, since there are many ways of discretizing the Langevin
equation
(\ref{lan}). If one sets $t_{i}=i\epsilon ,$ for $i=0, 1, 2,
\dots, M $,
$x_{i}=x_{t_{i}}$, and $F_{i}=F(x_{i})$, then the velocity $\dot{x}$ may
be approximated by $(x_{i}-x_{i-1})/\epsilon $. On the sliced time axis, the
force $F(x_{t})$ may act at any time within the slice $(t_{i},t_{i-1})$,
which is accounted for by a parameter $a$ and a discretization $F\rightarrow
aF_{i}+(1-a)F_{i-1}$.
Note that the discretized Langevin equation is assumed to be causal,
meaning that given the initial value of the stochastic variable
$x_0$ and the noise configurations $\eta_0,\eta_1,...,\eta_{M-1}$,
the Langevin equation uniquely determines the configurations
of the stochastic variable at later time, $x_1,x_2,...,x_M$.
The simplest choice of the right-hand side of the Langevin equation
compatible with the causality is to set it equal to $\eta_{i-1}$.
In general, one can replace $\eta_{i-1}$ by
$\sum_1^{M}A_{j-1,i-1}\eta_{i-1}$
with $A$ being an orthogonal matrix, $A^TA=1$. The latter is just
an evidence of the symmetry of the stochastic process with the white
noise with respect to orthogonal transformations $\eta_t\rightarrow
(A\eta)_t$.

Some specific values of the interpretation parameter
$a$ have been favored in the
literature, with $a=0$ or $1/2$ corresponding to the so-called It\^{o}- or
Stratonovich-related interpretation of the stochastic process (\ref{lan}),
respectively \cite{klauder,gard}. In the time-sliced path integral, these
values correspond to a prepoint or midpoint sliced action \cite
{klauder,PI}. Emphasizing the $a$-dependence of the sliced action, we shall
denote it by $S_{a}^{H}$. This action is supersymmetric for any $a$: $%
Q^HS_{a}^{H}=0$, and the sliced generator $Q^{H}=\sum_{i}(ic_{i}\partial
_{x_{i}}+p_{i}\partial _{\bar{c}_{i}})$ turns out to be independent of both
the interpretation parameter $a$ and the width $\epsilon $ of time slicing
\cite{klauder}. A shift of $a$ changes the action by the $Q$-exact term,
\begin{equation}
S_{a+\delta a}^H=S_{a}^H+\delta a\,Q^{H}\,G,  \label{da}
\end{equation}
where $G$ is a function of $a$ and a functional of $p,x,\bar{c},c$. This
makes the Ward identities independent of $a$, i.e. on the interpretation of
the Langevin equation. Indeed, setting $\delta a=-a$ we find $%
e^{-S_{a}^{H}}=e^{-S_{0}^{H}}e^{aQ^{H}\Phi }\equiv
e^{-S_{0}^{H}}(1+Q^{H}\Phi _{a}^{\prime })$. Substituting this relation into
(\ref{wi}) we observe that the $a$-dependence drops out from the Ward
identities because of the supersymmetry $Q^{H}S_{0}^{H}=0$ and the
nilpotency $(Q^{H})^{2}=0$.

The simplest situation arises for the It\^{o} choice, $a=0$. Then the
sliced fermion determinant $\Delta $ becomes a trivial constant independent
of $x$. In the continuum limit of the path integral, however, this choice is
inconvenient since then the limiting action $S_{0}$ {\em cannot\/} be
treated as an ordinary time integral over the continuum Lagrangian. Instead,
$S_{a=0}$ goes over into a so-called It\^{o} stochastic integral \cite
{klauder}. The It\^{o} integral calculus~\cite{gard} differs in several
respects form the ordinary one, most prominently by the property $\int
dx\neq \int dt\dot{x}$. This difficulty is avoided taking the Stratonovich
value $a=1/2$, for which the continuum limit of $S_{1/2}$ {\em is\/} an
ordinary integral \cite{klauder,PIM}. Splitting (\ref{da}) as $%
S_{a}=S_{1/2}+(a-1/2)Q\,G$, the non-Stratonovich part
vanishes in the continuum
limit because $Q$ does not depend on the slicing parameter $\epsilon $,
whereas $G$ is proportional to $\epsilon \rightarrow 0$ \cite{klauder}. For $%
a=1/2$, formula (\ref{zH}) has a conventional continuous interpretation as a
sum over paths, and can be treated by standard rules of path
integration
 (e.g., perturbation expansion around  Gaussian measures).
The price for this is the additional fermion interaction, which possesses as
an attractive feature the additional supersymmetry.

The aim of our work is to extend this supersymmetric path integral
representation to stochastic processes with higher time derivatives
\begin{equation}
L_{t}=\gamma (\partial _{t})\dot{x}_{t}+F(x_{t})=\eta _{t}\ ,  \label{hol}
\end{equation}
where $\gamma $ is a polynomial of any order $N-1$, thus producing $N$ time
derivatives on $x_{t}$. This Langevin equation may account for inertia via a
term $m\,\partial _{t}$ in $\gamma (\partial _{t})$, and/or an arbitrary
nonlocal friction $\int d\tau \gamma _{\tau }\dot{x}_{t-\tau
}=\sum_{n=0}^{N-1}\gamma _{n}\partial _{t}^{+1}{x}_{t}$ where $\gamma
_{n}=\int d\tau \gamma _{\tau }(-\tau )^{n}/n!$. The main problem is to find
a proper representation of the more complicated determinant $\Delta =\det
\delta _{x_{t^{\prime }}}L_{t}$ in terms of Grassmann variables. The
standard formula (\ref{det}), though formally applicable,
does not provide a proper representation of
the determinant of an
operator with higher-order derivatives
because of the boundary condition problem.
This problem is usually resolved via an operator representation
of the associated fermionic system.
In the stochastic context it has so far been discussed only for
the single time derivative \cite{klauder}. In the first-order
formalism, the fermion path
integral can be defined in terms of coherent states \cite{klauder}
with the above discussed vacuum-to-vacuum boundary conditions.
Higher-derivative theories, however, have many unphysical features, in
particular states with negative norms \cite{PU,brst}, and it is {\em a priori%
} unclear how to define the boundary conditions for the associated fermionic
path integral. In gauge theories, the Faddeev-Popov ghosts
give an example of a fermionic theory with higher-(second-)order derivatives.
There, unphysical consequences of the negative norms
of the ghost states
are avoided by imposing the so called BRST
invariant boundary conditions upon the path integral. For the above
stochastic determinant with
higher-order derivatives,
the correct boundary condition
are unknown.

\vskip 0.2cm {\bf 2}. The solution proposed by us in this work is best
illustrated by first treating Kramers' process where one more time
derivative is present, accounting for particle inertia, i.e. where $%
\gamma(\partial_t) = \partial_t + \gamma $ for a unit mass $m\equiv 1$.
Omitting
the time subscript of the stochastic variables, for brevity, we replace the
stochastic differential equation (\ref{hol}) by two coupled first-order
equations
\begin{eqnarray}
L_v&= &\dot{v} +\gamma v + F(x) = \nu_v,  \label{v} \\
L_{x}&= &\dot{x} - v= \nu_{x},  \label{x}
\end{eqnarray}
There are now two independent noise variables, which fluctuate according to
the path integral
\begin{eqnarray}
\langle F[x,v]\rangle &=& {\int}{\cal D}\nu_{x}{\cal D}\nu_v F[x,v]
\nonumber \\
&&\times e^{-1/2\int dt [ \nu_v^2/2(1-\sigma) + (\dot{\nu}_{x} +
\gamma\nu_{x})^2 /2\sigma ]}\ .  \label{noise}
\end{eqnarray}
A parameter $\sigma$ regulates the average size of deviations of $\dot x$
from $v$ in Eq.~(\ref{x}). If we regard the basic noise correlation
functions as  functional matrices $(D_n)_{tt^{\prime}} = \langle
\nu_{nt}\nu_{nt^{\prime}}\rangle$ for $n=x,v$, which act on functions of
time as linear operators $D_n f_t = \int
dt^{\prime}\,(D_n)_{tt^{\prime}}f_{t^{\prime}}$, the noise
correlation functions associated with (\ref{noise}) are
\begin{equation}
D_v= 1-\sigma\ ,\ \ \ D_{x}= \sigma e^{-\gamma t} \left(-\partial_t^{-1}
e^{2\gamma t}\partial_t^{-1} \right)e^{-\gamma t}\ .  \label{dxv}
\end{equation}
Substituting (\ref{x}) into (\ref{v}) we find the two-derivative version of (%
\ref{hol}), $\ddot{x} +\gamma \dot{x} + F(x) = \eta_\sigma$, driven by the
combined noise
\begin{equation}
\eta_\sigma =\nu_v + \dot{\nu}_{x} + \gamma \nu_{x}.\   \label{cnoise}
\end{equation}
This noise is white for {\em any\/} choice of $\sigma$:
\begin{equation}
\langle \eta_{\sigma t}\rangle = 0\ , \ \ \ \langle \eta_{\sigma
t}\eta_{\sigma t^{\prime}}\rangle = \delta_{tt^{\prime}}\ .  \label{ccor}
\end{equation}
Let $x_t[\eta]$ be a solution of the original Langevin equation (\ref{hol}),
and $x_t[\eta_\sigma] $ a solution of the system (\ref{x}), (\ref{v}). The
property (\ref{ccor}) implies that $x_t[\eta_\sigma] $ has the same
correlation functions as $x_t[ \eta]$, for {\em any\/} $\sigma$, thus
describing the same stochastic process. The freedom in choosing $\sigma$
will later be used to make the effective supersymmetric action local in time.

Once we have transformed Kramers' process into a system of coupled
first-order Langevin equations (\ref{x}) and (\ref{v})
which is a trivial extension of
the first-order equation (\ref{lan}) to a matrix form, there
obviously exists
a path
integral representation
analogous to
(\ref{zH}).
It is for this reason
that we have introduced a two
noise variable and a fluctuating
relation between $\dot{x}$ and $v$ in Eq.~(\ref{x}).
There is a complication though
in that the noise $\nu _{x}$ is no longer white since $D_{x}$ is
nonlocal in time. However, as observed above
this does not affect the supersymmetry in the canonical
form (\ref{zH}) of the path integral since the supersymmetry generator
$Q^{H} $ does not depend on $D_{x}$ (in contrast to $Q$).

Thus, having established the supersymmetric path integral representation
of the equivalent first-order stochastic system, our strategy is to
integrate out all auxiliary variables we have introduced and, thereby,
derive the proper boundary conditions for the fermionic path integral
in the higher order stochastic processes as well as to construct the
supersymmetry generator in the initial configuration space.

To prepare the notation for the later generalization to a stochastic
differential equation with $N$ derivatives, we rename the variables $x$ and $%
v$ as $x_n$, with $\alpha=1,N$, and for the moment
$N=2$. Only the equation for $x_N$
contains the force $F=F(x_1)$. The other equation just establishes a
fluctuating
equality between $\dot x_{1}$ and $x_2$, the original process being
described by $x\equiv x_1$. Inserting the stochastic equations (\ref{v}) and
(\ref{x}) into the exponent of (\ref{noise}), we repeat the previous
procedure and, choosing midpoint slicing with $a=1/2$ \`a la Stratonovich,
we obtain the path integral representation of the generating functional
\begin{eqnarray}
Z[J] &=& {\int} {\cal D}p {\cal D}x {\cal D}\bar{c}{\cal D}c\, e^{-S^{H} +
i\int dt Jx} \ ,  \label{hoz} \\
S^{H} &=& \sum _{n=1}^2 {\int} dt \left({{\frac{1}{2}}}p_n D_n p_n -ip_n L_n
+ \bar{c}_n \delta_{x_ m}L_n c_ m\right)\ .\nonumber
\end{eqnarray}
The generator of supersymmetry is
\begin{equation}
Q^{H} = {{\sum_{n=1}^2}} {\int} dt(i\bar{c}_n \delta_{x_n } - p_n \delta_{%
\bar{c}_n })\ .  \label{qh}
\end{equation}
It is readily verified that $Q^{H}S^{H}=0$, using the fact that $\sum_{ kmn}%
\bar{c}_ m c_ k (\delta_{z_ k}\delta_{z_n }L_ m)c_n \sim \sum_n c_n ^2 = 0$
due to the Grassmann nature of $c_n $. Explicitly, the Fermi part of the
action $S^H$ reads
\begin{equation}
S_f \!=\! {\int} dt\left[ \bar{c}_{x}\dot{c}_{x} \!+ \!\bar{c}_v\dot{c}_v
-\bar{c}%
_{x}{c}_v \!+\! \bar{c}_v{c}_{x}F^{\prime}(x)\! +\! \gamma\bar{c}_v{c}_v\right]
{}.
\label{sf}
\end{equation}

The Gaussian path integral over momenta in (\ref{hoz}) has a meaning without
time slicing, and can be performed to recover the Lagrangian version of the
supersymmetric action
\begin{eqnarray}
{S}= {\sum_{n=1}^2} {\int} dt{{\ \frac{1}{2}}} L_n D^{-1}_n L_n + S_f\ .
\label{ls}
\end{eqnarray}
The associated generator of supersymmetry is obtained from (\ref{qh}) by
substituting into $Q^H$ the solutions of the Hamilton equations of motion $%
p_n = iD_n ^{-1} L_n $ which extremize $S^H$ ($\delta_{p_n } S^{H} =0$),
leading to
\begin{eqnarray}
\tilde{Q}= {\sum_{n=1}^2} {\int} dt(i{c}_n \delta_{x_n } - iD_n ^{-1}L_n
\delta_{\bar{c}_n })\ .  \label{qs}
\end{eqnarray}
The final step consists in integrating out the auxiliary variable $x_2=v$,
which
only ap
pears quadratically in the bosonic part of the action. Making use of
the explicit form of $D_n $ given in (\ref{ccor}), we obtain the Lagrangian
form of the supersymmetric action
\begin{equation}
S_\sigma = {\int} dt {{\ \frac{1}{2}}} L_t \left( 1 + {{\frac{1}{1-\sigma}}}
\partial_tD_{x}\partial_t\right) L_t + S_f\ ,  \label{s}
\end{equation}
where $L_t$ is now the left-hand side of the initial equation (\ref{hol}),
for the Kramers process at hand: $L_t = \ddot{x}_t + \gamma\dot{x}_t +
F(x_t) $. At this stage, the effective action is nonlocal in time. Now we
take advantage of the freedom in choosing the parameter $\sigma$. We go to
the limit $\sigma\rightarrow 0$, in which case $D_{x} \sim \sigma$ vanishes,
reducing the action to the local form
\begin{equation}
S = S_0 = {\int} dt {{\frac{1}{2}}} L_t^2 +S_f\ .  \label{la}
\end{equation}

To find the generator of supersymmetry in this representation, we omit $%
\delta_{x_N}\equiv \delta_v$ in (\ref{qs}), and replace $x_N\equiv v$ by the
solution of the equation of motion
\begin{equation}
\delta_v\tilde{S} = -D_{x}^{-1} L_{x} + {{\frac{1}{1-\sigma}}} (-\partial_t
+\gamma)L_v = 0\ .  \label{vem}
\end{equation}
In the limit $\sigma\rightarrow 0$, $D_{x}^{-1} \sim \sigma^{-1}$ diverges,
leading to an exact equality $L_x=v - \dot{x}=0$,
rather than the fluctuating one (%
\ref{x}). To take the limit $\sigma\rightarrow 0$ in the operator (\ref{qs}%
), one must first substitute (\ref{vem}) into the would-be singular term
$D_x^{-1}L_x$
in $\tilde{Q}$, and then take the limit. The supersymmetry generator
assumes the final form
\begin{equation}
Q = {\int} dt \left[i{c}_{x}\delta_{x} - i (-\partial_t + \gamma)L_t \delta_{%
\bar{c}_{x}} -i L_t \delta_{\bar{c}_v}\right]\ .  \label{qq}
\end{equation}

The action (\ref{la}) provides us with the desired supersymmetric
description of processes with second-order time derivatives. An important
feature of the supersymmetry generated by $Q$ is that the supermultiplet
contains one boson field and two fermion fields. The reason for this
 is, of
course, that a boson field with $N$ time derivatives in the action carries $%
N $ particles, each of which must have a supersymmetric fermionic partner.
The fermion degrees of freedom have the conventional first order action,
which permits us to impose the vacuum-to-vacuum boundary conditions within the
coherent state representation of fermionic path integrals \cite{klauder}.
The boundary conditions for the bosonic path integral are the
causal ones: $x_{t=0} =
x_0$ and $\dot{x}_{t=0} = v_0 =\dot{x}_0$.

We have circumvented the problem of the boundary condition for the
determinant of a higher-order operator by enlarging the number of
Fermi fields,
thereby reducing the problem to the known one for the determinant of the
single-derivative
operator. What happens if we integrate out
the
auxiliary Grassmann variables $\bar{c}_{v},\ c_v$.
In these variables, the action
(\ref{sf}) is harmonic, driven by external
forces $\bar{c}_{x}$ and $c_{x}F^{\prime }(x)$.
After a quadratic completion
the integration
with the vacuum-to-vacuum
boundary condition yields $\det (\partial _{t}+\gamma )$.
The effective action for the other fermion pair becomes non-local
\begin{equation}
S_{f}=\int dt\left[ \bar{c}_{x}\dot{c}_{x}+\bar{c}_{x}(\partial _{t}+\gamma
)^{-1}(F^{\prime }(x)c_{x})\right] \ .  \label{sf1}
\end{equation}
The total effective action $S=S_{b}+S_{f}$ is still supersymmetric. The
supersymmetry is generated by the operator (\ref{qq}), if
the last term in $Q$ is dropped.
 The action (\ref{sf1}) is the first-order action. So with
the vacuum-to-vacuum boundary condition the integral over $\bar{c}%
_{x},c_{x}$ would also give a determinant. Thus we get the representation
\begin{equation}
\Delta =\det [\partial _{t}+\gamma ]\det [\partial _{t}+(\partial +\gamma
)^{-1}F^{\prime }(x)]\ .  \label{detdet}
\end{equation}
Invoking the formula for the determinant of a block matrix, the
non-locality in the second determinant can be removed, while
maintaining the linearity in the time derivative
\begin{eqnarray}
\Delta =\det
\left(
\begin{array}{cc}
\partial_t +\gamma & F^\prime \\
-1 & \partial_t
\end{array}
\right) ,
\label{2x2}
\end{eqnarray}
which is exactly the determinant arising from the two-noise process
(\ref{v}), (\ref{x}).
In this way we have
represented the determinant of the
second-order
operator as a determinant of a first-order operator
acting upon a higher-dimensional space
for which
the boundary conditions are known.

Thus, with the help of two coupled equations
driven by auxiliary noises we have succeeded
in giving a unique
meaning to the path integral representation
of the Kramers process.
The final path integral can be time-sliced
in any desired way (prepoint, postpoint, midpoint, or any combination of
these)---as long as the slicing
is done equally
in the bosonic and the fermionic
actions.
In Section 4,
the procedure will be generalized
to a
 friction coefficient
$\gamma $
which is a
function of $x$.

\vskip 0.2cm {\bf 3}. We now generalize
our construction to
stochastic processes of an arbitrary order $N$.
As a result we shall arrive at a
supersymmetric extension of general higher order Lagrangian systems
with a supermultiplet of $N$ fermion fields
which all
possess a good quantum theory due to their
first-order dynamics.

Consider a system of coupled stochastic processes
\begin{eqnarray}
L_{N} &=&\dot{x}_{N}+{{\sum_{n=1}^{N}}}\gamma _{n-1}x_{n}+F(x_{1})=\nu _{N}\
;  \label{holn} \\
L_{n} &=&\dot{x}_{n}-x_{n+1}=\nu _{n}\ ,\ \ \ n=N-1,N-2,...,1\ ,  \label{ln}
\end{eqnarray}
where $x_{1}\equiv x$. This stochastic process is equivalent to the original
one if we assume the noise average as being taken with the weight $%
e^{-S_{\nu }}$, generalizing that in (\ref{noise}) to
\begin{equation}
S_{\nu }={{\frac{1}{2}\int }}dt\left[ {{\frac{1}{1-\sigma }}}\nu _{N}^{2}+{{%
\sum_{n=1}^{N-1}\frac{1}{\sigma _{n}}}}(\Lambda _{N-n}\nu _{n})^{2}\right] \
,  \label{mnoise}
\end{equation}
where $\sigma =\sum_{n=1}^{N-1}\sigma _{n}$ and $\Lambda
_{n}=\sum_{m=0}^{n}\gamma _{N-m}\partial _{t}^{n-m},\ \gamma _{N}\equiv 1$.
As for $N=2$, equations (\ref{holn}) and (\ref{ln}) can b
e combined into a
single equation $L_{t}=\nu _{Nt}+\sum_{n=1}^{N-1}\Lambda _{N-n}\nu
_{nt}\equiv \eta _{\sigma t}$. {}From (\ref{mnoise}) follows that $\langle
\eta
_{\sigma t}\rangle=0$ and $\langle \eta _{\sigma t}\eta _{\sigma t^{\prime
}}\rangle=\delta _{tt^{\prime }}$.
Thus
the correlation functions of the system (\ref{holn}) are the same
as of the original one. Note also that
the combined noise correlation
functions do not depend on the parameters $\sigma _{n}$. We shall assign
some specific values to the $\sigma $'s to simplify the sequel formalism.

The Hamiltonian path integral for the stochastic system (\ref{holn}) and (%
\ref{ln}) has the form (\ref{hoz}), where the label $n$ runs
now from $1$ to $N$. With the same extension of the index sum, the operator $%
Q^{H}$ in (\ref{qh}) generates supersymmetry. The noise correlation
functions (\ref{dxv}) are generalized to $D_N = 1-\sum_{n=1}^{N-1}\sigma_n $
and $D_n = \sigma_n (\Lambda_{N-n} ^\dagger\Lambda_{N-n} )^{-1}$. After
integrating out the momenta $p_n $ we arrive at the action (\ref{ls}) with
the extended sum, and the generator of supersymmetry assumes the form (\ref
{qs}) with the extended sum.

Integrating out the auxiliary variables $x_{n}$ is now technically more
involved, but the integral is still Gaussian. A successive integration is
possible by observing that the fermion action does not depend on the
variables $x_{n}$ for $n>1$, the stochastic process being nonlinear only in
the physical variable $x_{1}\equiv x$. The classical equations of motion $%
\delta _{x_{n}}\tilde{S}=0$ can be written in the form
\begin{equation}
-D_{n-1}^{-1}L_{n-1}-\partial _{t}D_{n}^{-1}L_{n}+\gamma
_{n-1}D_{N}^{-1}L_{N}=0\ ,
\end{equation}
for $n=2,3,...,N$.
Combining the equations for $n=N$ and $n=N-1$, and the result with the
equation for $n=N-2$, and so on, we derive the relation
\begin{equation}
D_{n}^{-1}L_{n}={{\frac{1}{1-\sigma }}}\left[ {{\sum_{k=0}^{N-n}}}%
(-1)^{k}\gamma _{n+k}\partial _{t}^{k}\right] L_{N}\ ,
\label{dl}
\end{equation}
having inserted
$D_{N}=1-\sigma $ and with  $n=2,3,...,N$.
These expressions may be substituted into
the action (\ref{ls}), and the generator (\ref{qs}). As in the case $N=2$,
the supersymmetric Lagrangian action and the operator $Q$ turn out to have a
smooth limit $\sigma _{n}\rightarrow 0$. Since $D_{n}\sim 1/\sigma _{n}$, we
see from (\ref{dl}) that $L_{n}\rightarrow 0$, and we recover the physical
relations $\dot{x}_{n}=x_{n+1}$ and, hence, $x_{n}=\partial _{t}^{n}x$. The
action assumes the form (\ref{la}), with $L_{t}$ of Eq.~(\ref{hol}). The
generator of supersymmetry becomes
\begin{equation}
Q=i\int dt\left\{ c_{1}\delta _{x}-\sum\limits_{n=1}^{N}\left[
\sum\limits_{k=0}^{N-n}(-1)^{k}\gamma _{n+k}\partial _{t}^{k}L_{t}\right]
\,\delta _{\bar{c}_{n}}\right\} \ .  \label{mq}
\end{equation}
For convenience, we give the fermion action explicitly:
\begin{eqnarray}
S_{f} &=&{\int }dt\left[ {{\sum_{n=1}^{N}}}\bar{c}_{n}\dot{c}_{n}-\bar{c}%
_{n}c_{n+1}\right.  \nonumber \\
&&\left. ~~~~~~~~~+c_{N}\left( {{\sum_{n=0}^{N}}}\gamma
_{n-1}c_{n}+F^{\prime }(x)c_{1}\right) \right] .  \label{fa}
\end{eqnarray}
The operator (\ref{mq}) transforms the original stochastic variable $x=x_{1}$
into the Grassmann variable $c_{1}$, $Qx=ic_{1}$, whereas all the fermionic
variables are transformed into some functions of the only bosonic variable $%
x $. The fermionic action (\ref{fa}) is constructed in such a way that $%
QS_{f}$ depends only on $c_{1}$. The terms containing the other Grassmann
variables are cancelled amongst each other. The $c_{1}$-term is cancelled
against the term resulting from $QS_{b}$, i.e. $Q(S_{b}+S_{f})=0$. It is
important to realize that the fermions are coupled with each other, and thus
belong to an irreducible supermultiplet. The number of fermion is equal to
the highest order of the time derivative entering the bosonic action, as
observed before for $N=2$.

\vskip 0.2cm {\bf 4}. The idea of splitting the higher order Langevin
equation into a system of coupled first-order stochastic processes with a
combined noise can also be applied to construct a supersymmetric quantum
theory associated with the higher order stochastic process where the
coefficients $\gamma _{n}$ are functions of $x_{t}$. We illustrate this with
the example of Kramers' process with the friction coefficient being a
function of the stochastic variable $x_{t}$.

A straightforward replacement of $\gamma $ by $\gamma (x)$ in (\ref{v})
would yield a problem because the combined noise $\eta _{\sigma }$ appears
to be a function of $x_{t}$, making the system (\ref{v}), (\ref{x})
inequivalent to the original stochastic process (if the Gaussian
distributions for the auxiliary noises are assumed). To resolve this
problem, we take two coupled non-linear first-order processes
\begin{eqnarray}
L_{v} &=&\dot{v}+v+\lambda _{v}(x)=\nu _{\sigma }\ ,  \label{nv} \\
L_{x} &=&\dot{x}-v+\lambda _{x}(x)=\nu _{\sigma }\ .  \label{nx}
\end{eqnarray}
The functions $\lambda _{x,v}$ are subject to the condition
\begin{equation}
\lambda _{x}^{\prime }=\gamma -1\ ,\ \ \ \lambda _{v}=F-\lambda _{x}\ .
\label{ll}
\end{equation}
With the noise average defined by (\ref{noise}) and the condition (\ref{ll}%
), the stochastic system (\ref{nv}), (\ref{nx}) is equivalent to the
original system $L_{t}=\ddot{x}+\gamma (x)\dot{x}+F(x)=\eta $.

The difference between (\ref{nx}) and (\ref{x}) is just the extra force $%
\lambda _{x}$, which does not affect the derivation of the associated
supersymmetric action. Repeating calculations of section 2, we arrive at the
supersymmetric action $S=S_{b}+S_{f}$ where
\begin{eqnarray}
\!\!\!S_{b}\!\! &=&\!\!\frac{1}{2}\int dt\left( \ddot{x}+\gamma (x)\dot{x}%
+F(x)\right) ^{2}\ ,  \label{nsb} \\
\!\!\!S_{f}\!\! &=&\!\!\int dt\left[ \bar{c}_{x}\dot{c}_{x}+\bar{c}_{v}\dot{c%
}_{v}+\bar{c}_{x}c_{x}(\gamma (x)-1)+\bar{c}_{v}c_{v}\right.  \nonumber \\
&&~~~~~\left. -\bar{c}_{v}c_{x}(F^{\prime }(x)-\gamma (x)+1)-\bar{c}%
_{x}c_{v}\right] .  \label{nsf}
\end{eqnarray}
The supersymmetry generator has the form
\begin{equation}
Q=\int dt\left( ic_{x}\delta _{x}-iL_{t}\delta
_{\bar{c}_{v}}-i(-\partial
_{t}+1)L_{t}\delta _{\bar{c}_{x}}\right) \ .  \label{nq}
\end{equation}
It is not hard to verify that $QS=0$.

If we set $\gamma $ to be independent of $x$ in (\ref{nsf}), the fermionic
action does not turn into (\ref{sf}), in contrast to what one might expect.
The reason is that the fermionic path integral exhibits a large symmetry
associated with general canonical transformations on the Grassmann phase
space spanned by $\bar{c}$ and $c$. Recall that under canonical
transformations the canonical one-form $\sum_{n}\bar{c}_{n}dc_{n}$ is
invariant up to a total differential $dF(\bar{c},c)$. Also
the measure $%
\prod_{n}d\bar{c}_{n}dc_{n}$ remains unchanged.
Thus there exists
infinitely many equivalent supersymmetric representations of the same
stochastic process. The situation is similar to the BRST symmetry \cite{brst}
in gauge theories where the BRST charge is defined up to a general
canonical transformation. This freedom  can be used to simplify the
fermionic action or the Fermi-part of the supersymmetry generator.

This formal
invariance of the continuum phase-space path integral measure with respect
to canonical transformations
has been studied thoroughly
\cite{klauder2}
for bosonic phase spaces. A regularization of the continuum
phase-space path integral measure with respect to canonical
transformations on a phase space which is a Grassmann manifold
is still an open problem.

\vskip 0.3cm
\noindent {\bf Acknowledgment}:\newline
The authors are grateful to Drs. Glenn Barnich and Axel Pelster for many useful
discussions, and  to Prof. John Klauder
for comments.


\begin{references}
\bibitem{parisi}  G. Parisi and N. Sourlas, Phys.Rev.Lett. {\bf 43}, 744
(1979); Nucl.Phys. {\bf B206}, 321 (1982);\newline
M.V. Feigel'man and A.M. Tsvelik,
Sov.Phys. JETP, {\bf 56}, 823 (1982); Phys.Lett. {\bf 95A}, 469
(1983);\newline
 For
a comprehensive review see J. Zinn-Justin, {\em Quantum Field Theory and
Critical Phenomena} (2nd Edition, Clarendon Press, Oxford, 1993).

\bibitem{klauder}  H. Ezawa and J.
R. Klauder, Prog.Thor.Phys. {\bf 74}, 104
(1985);\newline
L.P. Singh and F. Steiner, Phys.Lett. {\bf 166B}, 155 (1986);\newline
H. Nakazato, K. Okano, L. Sch\"ulke and Y. Yamahaka, Nucl.Phys.
{\bf B346}, 611 (1990).

\bibitem{gard}  {C.W.~Gardiner}, {\em Handbook of Stochastic Methods\/}
(Springer Series in Synergetics, Vol. 13, Springer, Berlin, 1983).

\bibitem{PI}  H. Kleinert, {\it Path Integrals in Quantum Mechanics,
Statistics and Polymer Physics}, World Scientific, Second Edition, 1995.

\bibitem{PIM}  In Section 10.5 of Ref.~\cite{PI} it is shown that the
correct time slicing of an interaction $\int dt\,\dot{q}F(q)$ in a path
integral is of the midpoint type, corresponding to $a=1/2$. Sometimes this
is referred to as the {\em midpoint prescription\/} for defining the sliced
action, but it can actually be {\em derived\/} from the short-time action
along a classical orbit.

\bibitem{PU}  A. Pais and G.E. Uhlenbeck, Phys. Rev. {\bf 79}, 145
  (1950); \newline
M.V. Ostrogradsky, Mem. Acad. Sci. St-Petersburg, {\bf 6}, 385 (1850);
See also: E.T. Whittaker, {\em A Treatise on the Analytical Dynamics
of Particles and Rigid Bodies} (Cambridge University Press,
Cambridge, 1959);\newline
and Section 17.3 in Vol. II of H. Kleinert, {\it Gauge
Fields in Condensed Matter\/}, (World Scientific, Singapore, 1989).

\bibitem{brst}  M. Henneaux and C. Teitelboim, {\em Quantization of Gauge
Systems} (Princeton University Press, Princeton, 1992).

\bibitem{klauder2}  J.R. Klauder, Ann. Phys. (NY), {\bf 188}, 120 (1988).
\end{references}
\end{document}